\documentclass[aps,prl,twocolumn,showpacs,superscriptaddress]{revtex4}
\usepackage{amssymb,amsmath}
\usepackage{graphicx}
\usepackage{dcolumn}
\usepackage{bm}

\begin{document}
\title{First-Principles Modeling of Pt/LaAlO$_{3}$/SrTiO$_{3}$ Capacitors\\
              Under an External Bias Potential}
\author{Claudio Cazorla}
\affiliation{Institut de Ci$\grave{e}$ncia de Materials de Barcelona
            (ICMAB-CSIC), 08193 Bellaterra, Spain}
\author{Massimiliano Stengel}
\affiliation{Institut de Ci$\grave{e}$ncia de Materials de Barcelona 
            (ICMAB-CSIC), 08193 Bellaterra, Spain}
\affiliation{ICREA - Instituci\'o Catalana de Recerca i Estudis Avan\c{c}ats, 08010 Barcelona, Spain}
\email{mstengel@icmab.es}
\begin{abstract}
We study the electrical properties of Pt/LaAlO$_{3}$/SrTiO$_{3}$ 
capacitors under the action of an external bias potential, using 
first-principles simulations performed at constrained electric displacement field.
A complete set of band diagrams, together with the relevant electrical
characteristics (capacitance and \emph{built-in} fields),
are determined as a function of LaAlO$_{3}$ thickness and the applied potential. 
We find that the internal field in LaAlO$_3$ 
monotonically decreases with increasing thickness; hence, the 
occurrence of spontaneous Zener tunneling is ruled out in this 
system.
We discuss the implications of our results 
in the light of recent experimental observations on biased 
LaAlO$_{3}$/SrTiO$_{3}$ junctions involving metallic top electrodes. 
\end{abstract}
\pacs{73.20.-r, 71.15.-m, 71.70.-d}
\maketitle

\begin{figure}[t]
\centerline{
\includegraphics[width=3.2in,trim=0mm 100mm 0mm 10mm, clip]{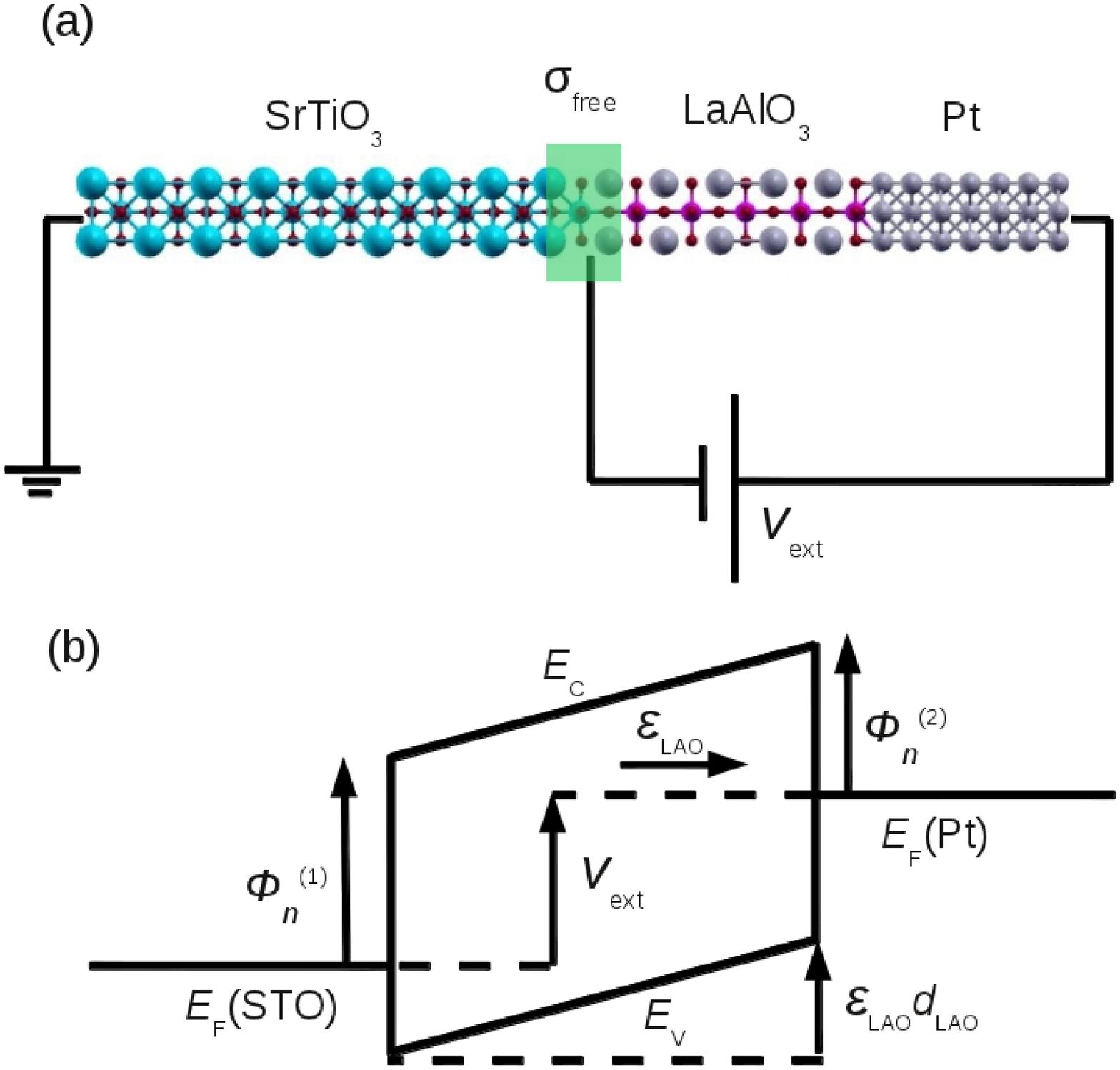}}
\vspace{-0.50cm}
\caption{(a)~Schematic illustration of the Pt/LAO/STO capacitor;    
large blue spheres represent Sr atoms, small blue Ti, 
red O, large purple La, pink Al and small purple Pt.
(b)~Pt/LAO/STO band diagram;
$\phi^{(1,2)}_{n}$ are the $n$-type SBH at the LAO/STO and Pt/LAO
junctions, $\mathcal{E}_{\rm LAO}$ is the electric field in LAO, 
$d_{\rm LAO}$ is the thickness of the film, 
$E_{\rm V}$ / $E_{\rm C}$ are the valence / conduction 
band edges, $E_{\rm F}$ are the Fermi levels, and $V_{\rm ext}$
is the external bias potential.}
\label{fig1}
\end{figure}

In recent years oxide-oxide heterojunctions have generated
widespread interest, because of their potential for applications 
in microelectronics and intriguing fundamental physics.
When thin films of polar LaAlO$_{3}$~(LAO), composed of formally 
charged (LaO)$^{+}$ and (AlO$_{2}$)$^{-}$ layers, are stacked on 
top of the non-polar $(001)$ surface of SrTiO$_{3}$~(STO), a 
two-dimensional metallic electron gas appears at the 
interface under suitable experimental conditions~\cite{ohtomo04}.
In particular, TiO$_{2}$-terminated LAO/STO interfaces are found to 
undergo an insulating-to-conducting transition (named
``electronic reconstruction'') when the 
thickness of the LAO film exceeds $3-4$ layers~\cite{thiel06}.
Alternatively, 2D metallicity can be switched on and off 
by applying an external bias potential between 
the conducting LAO/STO interface and a metallic electrode 
placed on top of the free LAO 
surface~\cite{Cen-08,Lu-11,Jany-10,cheng11,singh11,mannhart10}. 
In the latter setup (i.e. an asymmetric capacitor where the bottom 
electrode is the 2D electron gas, see Fig.~\ref{fig1}) novel striking 
phenomena have been observed recently, which hold 
promise for the realization of novel field-effect devices with 
high operation speed and low power consumption.

First, Singh-Bhalla {\em et al.}~\cite{singh11} reported a non-trivial 
dependence of the tunneling current on the LAO thickness, $d_{\rm LAO}$, 
with an abrupt increase at $d_{\rm LAO}=20$ unit cells. 
This was ascribed to the presence of a \emph{built-in} electric field,
which would cause Zener breakdown in the LAO film after exceeding
a certain $d_{\rm LAO}$ critical value. 
The electrical response of the tunnel junction also displayed a 
clear hysteretic behavior as a function of applied voltage, suggesting
interesting opportunities for memory applications.
Second, the measured capacitance was found to undergo a 
remarkable increase at low carrier densities~\cite{mannhart10}.
This enhancement was ascribed to the peculiar electronic properties of 
the 2D electron gas at the LAO/STO interface, and in particular
to a ``negative compressibility'' regime. 
While such a behavior was already known in the context of semiconductor
heterojunctions, its unprecedented magnitude (40\%) in the LAO/STO 
system challenges the current theoretical understanding of this effect.
Rationalizing these phenomena in terms of the microscopic properties
of the LAO/STO and LAO/electrode junctions is very desirable in sight 
of future progress.
In this context, first-principles electronic structure methods appear
ideally suited to describing, with unbiased accuracy, the subtle interplay 
between carrier confinement, polar distortions and electrical perturbations 
applied to the sample. Indeed the LAO/STO system has been addressed by a 
large number of ab-initio studies in the past few 
years~\cite{pentcheva09,janicka09,delugas11,popovic08,chambers11,chen10,son09,
bristowe11}.
However, studying the phenomena described in
Refs.~\cite{singh11} and~\cite{mannhart10}
entails some additional technical challenges, because of 
the necessity of introducing an external voltage in a complex 
capacitor system that is overall metallic.  

\begin{figure}[t]
\centerline{
\includegraphics[width=1.00\linewidth,trim=0mm 40mm 0mm 0mm,clip]{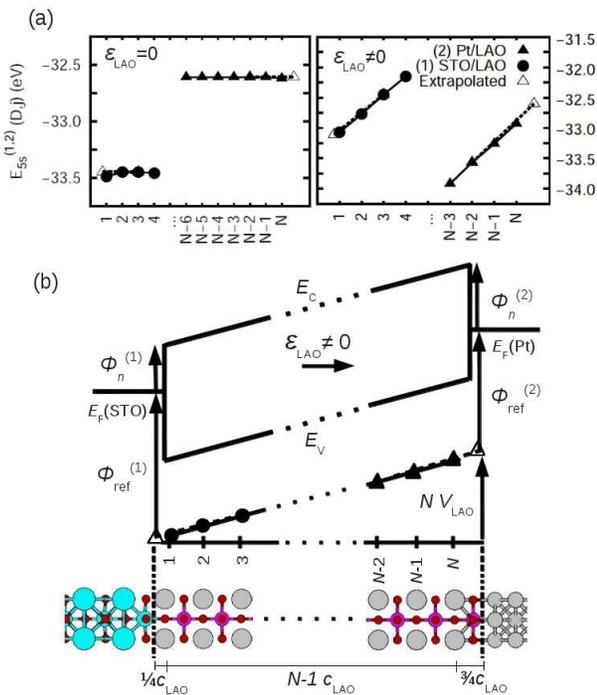}}
\vspace{-0.55cm}
\caption{
(a)~Layer-resolved La~$5s$ semicore energies, $E_{\rm 5s}^{(1,2)}$,
obtained at $D = -0.5 \, \, e/S$~(left) and $-0.3 \, \, e/S$~(right). 
Triangles represent the extrapolated values at the interfaces,
$\phi_{\rm ref}^{(1,2)}$.  
(b)~Sketch of the Pt/LAO/STO band diagram and details
of the extrapolation scheme (see text).}
\label{fig2}
\end{figure}

In this Letter, we propose an elegant solution to this problem
by using recently developed methods~\cite{stengel09,stengel11}
to perform \emph{ab initio} simulations at constant electric 
displacement $D$.
In particular, we show that: (i)~the internal electric
field ($\mathcal{E}_{\rm LAO}$) in a Pt/LAO/STO capacitor is 
not an intrinsic and fixed quantity, but a function of both applied bias 
and LAO thickness; (ii) at zero bias, $\mathcal{E}_{\rm LAO}$ 
monotonically decreases with film thickness as $\sim 1/d_{\rm LAO}$,
so that no Zener breakdown ever occurs; 
(iii) the LAO/STO interface is associated with a capacitance, 
predominantly due to band bending effects, which is constant within 
a wide range of carrier densities.
Based on these findings we deduce a complete set 
of \emph{ab initio} band diagrams and a simple analytical expression 
for $\mathcal{E}_{\rm LAO}$ as a function of the relevant parameters.
Finally, we discuss the implications of our results for the interpretation
of experimental observations in~\cite{mannhart10} and~\cite{singh11}. 

To start with, we are interested in determining the 
electronic and structural properties of an arbitrarily 
thick Pt/LAO/STO capacitor subjected to an external bias 
potential $V_{\rm ext}$.
Instead of working directly at fixed $V_{\rm ext}$, we
use the electric displacement field $D$ as independent electrical 
variable; 
this has clear advantages from the point of view of
modeling~\cite{wu08}, as it allows one to break down a layered
system into smaller constituents and treat them separately 
(``locality principle''). 
(Note that working at fixed $V$ or at fixed $D$ yields the exact same
information, except that we express it in terms of two different
macroscopic electrical variables; these are related by a Legendre
transformation~\cite{stengel09}.)
In our case, we shall consider the LAO/STO 
interface [(1) henceforth], bulk LAO and the Pt/LAO 
interface [(2) henceforth].
Following the arguments of Refs.~[\onlinecite{stengel09}] and~[\onlinecite{stengel09c}], 
we can decompose the total potential drop $V_{\rm tot}$ across the 
capacitor into three terms~(see Fig.~\ref{fig1}),
\begin{equation} 
V_{\rm tot}(D , N) = \phi_{n}^{(1)}(D) - \phi_{n}^{(2)}(D) + N \, V_{\rm LAO}(D)~.
\label{eq4}
\end{equation}
Here $V_{\rm LAO} = -\mathcal{E}_{\rm LAO} \, c_{\rm LAO}$ ($c_{\rm LAO}$ is the 
out-of-plane lattice parameter) is the potential drop across one  
unit cell of bulk LAO, $N$ the number of unit cells, 
and $\phi_{n}^{(1,2)}$ are the ($D$-dependent) $n$-type Schottky 
barrier heights (SBH) at the metal/insulator junctions~\cite{stengel09}.
Note that to obtain the ground state of the system at a given applied 
potential $V_{\rm ext}$ one simply needs to invert Eq.~(\ref{eq4}) and
solve for $V_{\rm tot}(D , N) = V_{\rm ext}$.
This way, the daunting problem of simulating the full Pt/LAO/STO 
capacitor under an external bias reduces to the more familiar task of 
calculating SBH at metal/insulator interfaces as a function of $D$.
As we shall explain in the following, it is relatively straightforward to 
do this with a standard first-principles code, without the
need for a specialized finite-field (or even Berry-phase) implementation.

Our calculations are performed within the local density approximation of
density functional theory and the PAW method~\cite{blochl94}, as implemented
in the in-house code {\sc Lautrec}~\cite{note0}.
We compute the quantities on the right hand side of Eq.~(\ref{eq4}) by means of 
three separate calculations: two interface models within a X/LAO/vacuum slab geometry, 
where X is the metallic electrode (either Pt or STO), and a periodic LAO bulk model.
For the X/LAO interfaces we use stacks of $10/7$ (X=Pt) and 
$12/5$ (X=STO) layers respectively. The in-plane periodicity is set to 
$1\times 1$ perovskite cells, with the lattice parameter fixed to the 
theoretical equilibrium value of bulk STO ($a_{\rm STO}=3.85$~\AA).
To constrain the electric displacement to a given value we introduce a layer of 
bound charges $Q$ at the free LAO surface via the virtual crystal 
approximation~\cite{note1}. By applying a dipole correction in  
vacuum, we enforce $D = 0$ outside the free surfaces; then, provided that the
surfaces remain locally insulating, we have $D=Q/S$~\cite{stengel11,stengel11b}.
We explore values of $D$ within the range $-0.5 \, \, e/S \, \le D_{\rm LAO} \le -0.3 \, \, e/S$ 
($D_{\rm STO}$ is set to zero). 

\begin{figure}[t]
\centerline{
\includegraphics[width=1.00\linewidth]{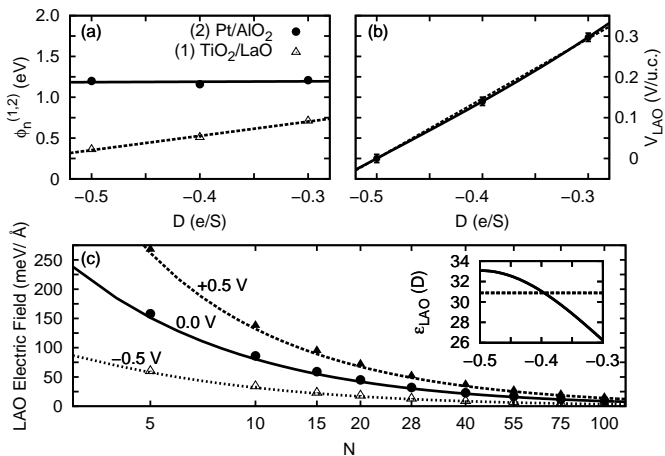}}
\vspace{-0.20cm}
\caption{
(a)~Calculated $n$-SBH as a function of $D$. 
The curves represent the linear fits
 $\phi_{n}^{(1)}(D) = 0.35 + 1.62 \, (D + e/2S)$ and
 $\phi_{n}^{(2)}(D) = 1.18 + 0.05 \, (D + e/2S)$ (in units of $V$).
(b)~Calculated potential drop per unit cell of bulk LAO; 
 $V_{\rm LAO}(D) = 1.27 \, (D + e/2S) + 2.24 \, (D + e/2S)^{3}$ 
 (solid line) and $V_{\rm LAO}(D) = 1.36 \, (D + e/2S)$ (dashed line).     
(c)~$\mathcal{E}_{\rm LAO}$ as a function of the number of LAO layers 
    and bias potential $-0.5 \le V_{\rm ext} \le 0.5$~V. The $x$-axis is in 
    logarithmic scale. Symbols and lines represent results obtained 
    with Eqs.~(\ref{eq4}) and~(\ref{eq6}), respectively. Inset:
    Static dielectric constant of bulk LAO as a function of $DS/e$ (solid curve) 
    and the corresponding averaged value $\overline{\epsilon}_{\rm LAO}$ (dashed line).}
\label{fig3}
\end{figure}

To extract the dependence of the SBH on $D$, we use computational 
techniques similar to those in Refs.~[\onlinecite{stengel09b}] and~[\onlinecite{stengel11b}]. 
In particular, for the estimation of the band offset $\phi_{n}^{(1,2)}(D)$
we need two independent quantities, the so-called \emph{lineup} 
term (an interface property) and the \emph{band-structure} 
term (a bulk property)~\cite{stengel11b}.
The lineup term relies on the choice
of a reference energy in the insulating LAO layer; here we use 
the La~$5s$ semicore energy, that we extract from the 
layer-resolved density of states of the slab models. We define as
$E_{5s}^{(1,2)}(D,j)$ the energy location of the La~$5s$ 
peak in the $j$-th LaO layer of the (1) or (2) system, 
referred to the respective Fermi level.
In the case of zero field, $E_{5s}^{(1,2)}(D,j)$ converges to
a constant value for layers $j$ lying sufficiently far away 
from the interface or the surface (typically two-three unit 
cells), and the definition of the lineup term is straightforward. 
Conversely, when the electric field in LAO is 
nonzero~(see Fig.~\ref{fig2}), $E_{5s}$ has a linear
dependence on the layer index. In particular, deep enough in
the film, the variation of $E_{5s}$ between two consecutive
cells corresponds precisely to the bulk internal field at the
same value of $D$,
\begin{equation}
E_{5s}^{(1,2)}(D,j+1)-E_{5s}^{(1,2)}(D,j) = V_{\rm LAO}(D).
\end{equation}
Therefore, once $V_{\rm LAO}(D)$ is known (from the bulk calculations), 
we can define an \emph{extrapolated} reference energy 
(see Fig.~\ref{fig2}),
\begin{equation}
E_{\rm ref}^{(1,2)}(D,j) = k(D) + j V_{\rm LAO}(D),
\end{equation}
where the constant $k(D)$ is chosen in such a way that 
$E_{\rm ref}(D,j)= E_{5s}(D,j)$ far from the interface.
Then, we define the lineup term as 
\begin{equation}
\phi_{\rm ref}^{(1,2)}(D) = E_{\rm ref}^{(1,2)}(D,j_{\rm I}^{(1,2)}),
\end{equation}
where $j_{\rm I}^{(1,2)}$ indicates the interface plane location; 
here we assume $j_{\rm I}^{(1)}=3/4$ and $j_{\rm I}^{(2)}=1/4$
(see Fig.~\ref{fig2}). 
Finally, to obtain the $n$-type SBH $\phi_{n}^{(1,2)}(D)$ we just need to 
add to $\phi_{\rm ref}^{(1,2)}(D)$ the $D$-dependent band-structure
term [we calculate it in the bulk, as the difference between the
La$(5s)$ level and the conduction band minimum], $E_{\rm CBM}(D)$,
\begin{equation}
\phi_{n}^{(1,2)}(D) = \phi_{\rm ref}^{(1,2)}(D) + E_{\rm CBM}(D).
\end{equation}

Our results for $\phi_{n}^{(1,2)}(D)$ and $V_{\rm LAO}(D)$ are 
shown in Fig.~\ref{fig3}(a,b).  We find that $\phi_{n}^{(1,2)}(D)$ 
behave linearly within the whole studied range. Consequently,  
the \emph{total} potential drop due to both interfaces can be expressed as 
$V_{\rm I}(D) = V_{\rm I}^{0} + C_{\rm I}^{-1} \, \left(D + e/2S \right)$,     
where $V_{\rm I}^{0}$ is the total interface potential drop at $D = -e/2S$ 
and the coefficient $C_{\rm I}^{-1}$ physically corresponds to the 
overall inverse interface capacitance density~\cite{stengel09,stengel09c}. 
Considering the potential drop in bulk LAO also to be linear 
in $D$~[see Fig.~\ref{fig3}(b)], i.e.
$V_{\rm LAO}(D) \approx C_{b}^{-1} \, \left(D + e/2S \right)$,
we obtain a simple yet very accurate analytical expression 
for the electric displacement field in LAO, 
\begin{equation} 
D \, (V_{\rm ext},N) + \frac{e}{2S} \approx  \frac{V_{\rm ext} - V_{\rm I}^{0}}{N \, C_{b}^{-1} + C_{\rm I}^{-1}}~,    
\label{eq6}
\end{equation} 
where $C_{b}^{-1} = 1.36$~m$^{2}$/F, $V_{\rm I}^{0} = -0.83$~V,
and $C_{\rm I}^{-1} = 1.57$~m$^{2}$/F ($S \equiv a_{\rm STO}^{2}$).    
The inverse bulk LAO capacitance density, $C_{b}^{-1}$, is directly related to 
the static dielectric constant of LAO, $\overline{\epsilon}_{\rm LAO}$, via 
$C_{b}^{-1} = c_{\rm LAO} \, / \, (\epsilon_0 \overline{\epsilon}_{\rm LAO})$~; 
in our computational model we have $\overline{\epsilon}_{\rm LAO} = 31$.
For a given value of $D$, we compute the corresponding electric field 
using $\mathcal{E}_{\rm LAO} = (D + e/2S) \, / \, \overline{\epsilon}_{\rm LAO}$. 
[In the ``exact'' treatment we replace the average $\overline{\epsilon}_{\rm LAO}$ 
with the calculated $\epsilon_{\rm LAO}(D)$, see the inset of Fig.~\ref{fig3}(c)]. 
Our results for $\mathcal{E}_{\rm LAO}(V_{\rm ext},N)$ 
are shown in Fig.~\ref{fig3}(c). It is worth noting that the outcomes of 
Eq.~(\ref{eq4})~[exact, considering non-linearities 
in $V_{\rm LAO}(D)$] and Eq.~(\ref{eq6}) (approximate) are in excellent 
agreement. 
In both cases we find that $\mathcal{E}_{\rm LAO}(V_{\rm ext},N)$ 
monotonically decreases with the inverse of LAO thickness, independently 
of the  applied bias potential.  

These results have important implications concerning the interpretation
of the experiments reported in Ref.~[\onlinecite{singh11}].
First, no intrinsic \emph{built-in} LAO electric 
field exists in the short-circuited Pt/LAO/STO capacitor
system, contrary to the assumptions of Singh-Bhalla {\em et al.}~\cite{singh11}.
Second, and most importantly, the uniform decrease of 
$\mathcal{E}_{\rm LAO}(V_{\rm ext},N)$ with LAO thickness
in practice rules out the hypothesis of Zener tunneling that
was used in Ref.~\cite{singh11} to explain the abrupt 
increase of the tunneling current at $N \sim 20$. 
Indeed, as $\mathcal{E}_{\rm LAO}(V_{\rm ext},N)$ decreases with
increasing $N$, Zener tunneling becomes increasingly less likely 
in thicker films; 
note that in our model the LAO valence band \emph{never} goes above the 
STO Fermi level unless a strong external potential $V_{\rm ext}\sim 1.8-1.9$ 
eV is applied.
An alternative interpretation of the aforementioned tunneling experiments 
rests on the results of Ref.~\cite{cancellieri11}. Precisely
at a thickness of $\sim 20$ unit cell, Cancellieri {\em et al.} reported
the onset of in-plane strain relaxation in the LAO overlayer.
Such relaxation processes are known to induce defects (e.g. cracks,
or misfit dislocations) in the LAO film~\cite{cancellieri11,wang08}. It is
not unreasonable to think that these defects might constitute 
preferential paths for electron or hole conduction; also, these 
defect-mediated conduction processes might explain the hysteretic
behavior of the electrical diagrams of Ref.~\cite{singh11}.
(The authors invoked the presence of switchable dipoles at the
LAO/STO interface, which appears puzzling as neither LAO nor
STO are ferroelectric.)

\begin{figure}[t]
\centerline{
\includegraphics[width=3.0in,clip]{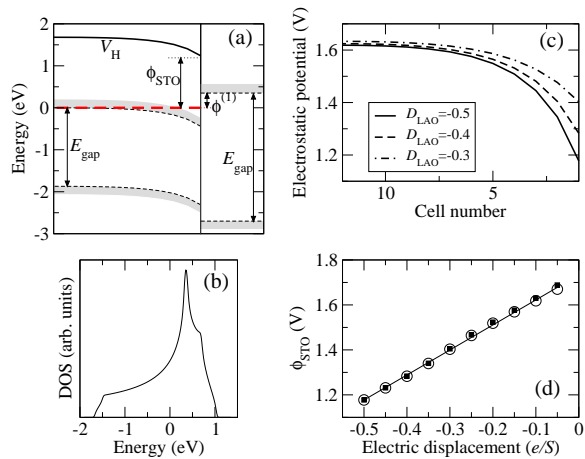}}%
\caption{
(a)~Mesoscopic band diagram of the LAO/STO 
junction. The electrostatic potential in STO, $V_{\rm H}$, 
(thick solid curve), the conduction and valence
band edges (shaded dashed curves) and the Fermi level (thick
dashed line) are shown. 
(b)~Ti $t_{2g}$ density of states of our TB model ($V_{\rm H}$ is
constant and set to zero).
(c)~$V_{\rm H}$ as calculated
with the TB model (Fermi levels are shifted to zero).
(d)~Electrostatic potential at the interfacial Ti site as a function 
of $D$. Empty circles~(filled squares) are calculated for 
a $12$~($24$)-layer STO slab and the solid line represents a 
linear fit to the data.}
\label{fig4}
\end{figure}
 
Finally, we shall comment on the microscopic mechanisms
that lead to the calculated interface properties.
The total interface capacitance density can be expressed as 
the sum of two contributions,  
\begin{equation}
C_{\rm I}^{-1} = \frac{d}{dD}\,(\phi_{n}^{(1)} - \phi_{n}^{(2)}) = \frac{1}{C^{(1)}} + \frac{1}{C^{(2)}}~,
\label{eq7}
\end{equation}
each due to a different interface.
The variation of $\phi_{n}^{(2)}(D)$ with 
$D$ is very small~(see Fig.~\ref{fig3});
this means that the Pt electrode behaves ideally (``perfect'' 
screening~\cite{stengel11b}), i.e. $1 / C^{(2)} \sim 0$, and 
therefore we restrict the following analysis to the LAO/STO junction. 
Specifically, we shall consider the effects of band-bending 
on $C_{(1)}^{-1}$, i.e. the $D$-dependent interfacial potential
drop that occurs in STO due to the presence of the confined charge 
carriers~\cite{yoshimatsu08}.
To estimate the band-bending contribution, $\phi_{\rm STO}(D)$,   
we have employed the tight-binding (TB) model~\cite{note2} introduced 
in Ref.~\cite{stengel11}. 
We define $\phi_{\rm STO}(D)$ as the value of the
electrostatic potential in STO at the interfacial Ti site, referred
to the Fermi level of the confined electron gas, at a given $D$~[see
Fig.~\ref{fig4}(a,c)].
Our results show that $\phi_{\rm STO} (D)$ behaves linearly 
within a wide range of values~[Fig.~\ref{fig4}(d)]. The band-bending 
contribution to $C_{(1)}^{-1}$ corresponds to the slope of the 
linear fit, which amounts to $1.03$~m$^{2}$/F. This contribution 
constitutes a significant fraction (approximately two thirds) of $C_{(1)}^{-1}$.
The remainder of $C_{(1)}^{-1}$ is likely to be caused by other effects
that typically occur at oxide-oxide interfaces, e.g. intrinsic
dielectric ``dead layers''~\cite{nature2006,copie09}. As these 
appear to be of secondary importance here, we have not
pursued this analysis further.

The above results are timely for the interpretation 
of recent capacitance measurements of thin-film 
electrode/LAO/STO heterostructures~\cite{mannhart10}.
In the regime of large carrier densities, Li \emph{et al.} measured 
a dielectric constant of the LAO film of $\epsilon_{\rm LAO} \sim 18$, 
which is significantly smaller than the typical value reported for LAO 
single crystals ($\epsilon_{\rm LAO} = 25-30$).
Our results indicate that this discrepancy may be related to
\emph{interfacial} effects, and in particular to the relatively
small value of $C^{(1)}$, rather than to the quality of the LAO
film. 
It is worth noting that Li \emph{et al.}~\cite{mannhart10} 
used YBCO electrodes, instead of the Pt electrodes considered here.
This implies that a possible contribution from the upper electrode
interface should be in principle taken into account (recall that 
$1/C^{(2)}$ essentially vanishes in the present Pt/LAO case) when
comparing our data to those of Ref.~\cite{mannhart10}.
In this context, note that Eq.~(\ref{eq6}) is completely general
and applies to any electrode/LAO/STO configuration; one simply needs
to replace two parameters (that are specific to the electrode/LAO
interface), the Schottky barrier at zero field $\phi_{n}^{(2)}$ and
the interface capacitance $C^{(2)}$.
In other words, one does not need to repeat the calculations 
of the LAO/STO interface, since we have essentially
decoupled the bottom from the top electrode and their properties
can be computed separately.

In summary, we have studied the electrical properties of
metal/LAO/STO capacitors fully from first-principles, determining a 
complete physical picture of the band offsets and internal fields.
Our results and methodologies open new avenues in the first-principles
study of functional oxide heterostructures, and provide
useful guidelines for the interpretation of the available experimental 
data on this system.

This work was supported by MICINN-Spain (Grants No. MAT2010-18113 and 
No. CSD2007-00041), ICREA, and the EC-FP7 project OxIDES (Grant No. CP-FP
228989-2). Computing time was kindly provided by BSC-RES and CESGA.


\begin{thebibliography}{30}
\bibitem{note1} The additional charge density is introduced by substituting a cation at the 
               ferroelectric surface by a fictitious cation of different formal valence. As
               we are interested in exploring intermediate values of $D$, we generated 
               fictitious pseudopotentials that reproduced fractional nuclear charges.
\bibitem{note2} This is based on an electronic $t_{2g}$-Hamiltonian describing the
                hoppings between Ti-centered orbitals, wherein the electrical boundary conditions 
                are defined by the electric displacement field in LAO.
\end{thebibliography}
\end{document}